# Efficient Monte Carlo Simulation of Biological Aging

T.J.P.Penna and D.Stauffer*
*Institute of Physics, Fluminense Federal University*
*C.P. 100296, 24001-970, Niterói, RJ, Brazil*

A bit-string model of biological life-histories is parallelized, with hundreds of millions of individuals. It gives the desired drastic decay of survival probabilities with increasing age for 32 age intervals.

*Present and permanent address: Institute for Theoretical Physics, Cologne University, D-50923 Köln, Germany

There are many theories why people get old: programmed cell death, oxygen radicals, mechanical wear and tear, etc. Rose[1] suggested that methods similar to mathematical physics should be applied to this problem, and several Monte Carlo studies of biological aging have been published recently[2]. They, as well as the present paper, try to explain aging through population dynamics and mutations: Bad mutations affecting young age before reproduction reduce population growth much stronger than if the same mutations affect only old age after reproduction. This Darwinistic selection pressure thus keeps youth more free from deleterious mutations than old age, and thus leads to a death rate increasing with age. The aim of the present work is to explain the algorithm for a numerically particularly efficient model[3] and to see if it gives biologically reasonable results (survival rate decaying rapidly at old age) for parameters which require large-scale parallelization. This model has the advantage, compared to earlier ones[2], that a large number of time intervals can easily be incorporated into the life history of an individual, in contrast to only two intervals (youth and adulthood).

We start with a large number of individuals having in their genes for each "year" $t$ of their lifes at most one mutation which becomes relevant in this year and the following years only: $t = 1, 2, \ldots, 32$. All mutations are deleterious (bad), and an individual dies if the number of mutations relevant at time $t$ reaches a certain limit $T$. These mutations are stored as bits in a 32-bit computer word defining the genome of the individual. They are hereditary in the sense that at the birth of a child they are given on to this child, except for $M$ randomly selected bits which are toggled at birth. These new mutations of rate $M$ affect only the child in its later life, not the parent; and these new mutations may not only increase but also decrease the number of old bad mutations stored in the computer word: If the randomly selected bit was already set to unity before (indicating an inherited disease) it is now reversed to zero such that no bad mutation is stored at that age. (Sexual reproduction is ignored at present and could be taken into account by a genetic algorithm mixing two bit strings.) In each year, with probability $m$, one child is produced by each surviving individual. Survival requires that the total number of set bits in positions $1, 2, \ldots, t$ of the computer word is less than $T$. In addition, we let individuals with a lower number of mutations survive only with the Verhulst probability $(1 - N/N_{max})$ where $N$ is the total number of invividua and $N_{max}$, typically ten times the initial $N$, is a maximum number of the population due to food and space restrictions.

Initially, the bits representing the genome are set randomly, half up and half down. The peculiar dangers of childhood are ignored here; thus babies of age $t = 0$ should be interpreted as offspring which has overcome these initial problems. Only individuals older than some minimum age get children.

This model is similar to that of Hötzel[4] in that more than two age intervals are taken into account. Hötzel simulated, for four aging steps, biased random walks on a large four-dimensional lattice. In this sense, the present new mutations happening at birth correspond to random walks on a 32-dimensional hypercube with $2^{32}$ sites. Our bit storage allows for a simple transfer of genetic information from parent to child: 32 age intervals in one operation only[5]. More biological background on mutations, though not for the purpose of aging, is given by Lynch and Gabriel[6] in connection with their Monte Carlo simulations.

To parallelize this algorithm on a multiple-data multiple-instruction computer with distributed memory is quite easy, since only through the food restriction factor $(1 - N/N_{max})$ the different nodes of a parallel computer are coupled. The total number $N$ can be easily evaluated by the global summation function `gisum` on the Intel Hypercube or Paragon computer. If the population in one of the nodes has died out, the loop over the predetermined number of iterations (years) still has to be continued since otherwise due to the synchronizing `gisum` all the other nodes would be blocked. A parallel computer with large total memory is required if, as is often the case, the population first decreases by several order of magnitude before Darwinistic selection of the fittest lets it grow again.

In the complete Fortran program listed as Fig.1, `minage` is the minimum age for reproduction, and our variables $T, N_{max}, N, M$ are denoted as `limit, nindm, npopul, mutr`; `ind` is the maximum number of individuals allowed because of memory limitations. The `data` statement gives initial parameters. Now we explain the various `do`-loops: Loop 1 produces isolated bits set at position $t = 1, 2, \ldots, 32$ and used later for the new mutations. Loops 2 and 3 generate the random initial genes `mgen`, using multiplication with 16807 for random number generation. The main loop 4 counts the iteration. Loop 5 over all individuals is realized by `goto 5` to allow for a fluctuating total number of individua. The bits (bad mutations) up to the current age are counted in loop 6. New mutations are introduced in loop 7 at birth. For parallelization we have to keep in memory the population for one node (loop 9) before all nodes are summed up by `call gisum`. At the end we print out the population as a function of age and the ratio between consecutive ages; the latter ones are the survival rates in a stationary state.

```
      parameter(ind=2000000)
      byte age(ind)
      real*8 dclock
      dimension mgen(ind),mbit(32),number(0:33)
     1     ,iwork(0:33),numt(0:33)
      data mutr/1/, limit/ 3/, iseed/1/, max/ 1000/
     1     , birth/0.11/, npopul/200000 /
     1     , minage/ 8/, nindm/20000000/
      time=dclock()
      node=mynode(i)
      if(node.eq.0)
     1 print 100, ind,npopul,mutr,limit,iseed,max,minage,nindm
     1 ,numnodes(),birth
```

```
      ibm=iseed*2-1+2*node
      ibirth=2147483647.0*(2.0*birth-1.0)
      mbit(1)=1
      number(0)=npopul
      do 1 j=1,32
      number(j)=0
 1    if(j.gt.1) mbit(j)=ishft(mbit(j-1),1)
      do 2 i=1,npopul
      ici=0
      do 3 j=1,32
      ici=ishft(ici,1)
      ibm=16807*ibm
 3    if(ibm.lt.0) ici=ici+1
      mgen(i)=ici
 2    age(i)=0
      do 4 itime=1,max
      if(npopul.le.0.or.npopul.gt.ind) goto 4
      iverhu=2147483647.0*(npopul*2.0/nindm-1.0)
      i=0
 5    i=i+1
10    number(age(i)) = number(age(i))-1
      if(age(i).le.32) age(i)=age(i)+1
      number(age(i)) = number(age(i))+1
      k=mgen(i)
      negmut=0
      do 6 j=1,age(i)
      negmut=negmut+iand(1,k)
 6    k=ishft(k,-1)
      ibm=ibm*16807
      if(ibm.lt.iverhu.or.negmut.ge.limit.or.age(i).eq.33) then
        number(age(i))=number(age(i))-1
        age (i)=age (npopul)
        mgen(i)=mgen(npopul)
        npopul=npopul-1
        goto 10
      end if
      ibm=ibm*65539
      if(ibm.lt.ibirth.and.age(i).gt.minage) then
        npopul=npopul+1
        if(npopul.gt.ind) goto 4
        mgen(npopul)=mgen(i)
        if(mutr.eq.0) goto 8
        do 7 k=1,mutr
```

```
        ibm=ibm*16807
 7      mgen(npopul)=ieor(mgen(i),mbit(1+ishft(ibm,-27)))
 8      age(npopul)=0
        number(0)=number(0)+1
      end if
      if(i.lt.npopul) goto 5
      if(itime.le.minage+32.or.itime.eq.(itime/minage)*minage) then
        npopt=npopul
        do 9 j=0,33
 9      numt(j)=number(j)
        call gisum(npopt,1,idummy)
        call gisum(numt,34,iwork)
        if (node.eq.0)
     1 print 100, itime,npopt,(numt(j),j=0,32,8),number(1)
      endif
 4    continue
100   format(1x,8i9,/,i9,f9.6)
      if (node.eq.0) print 102, numt
102   format(1x,10i7,/,1x,10i7,/,1x,14i5)
      if (node.eq.0) print 101, (numt(i)/(0.001+numt(i-1)),i=1,32)
101   format(1x,13f6.3,/,1x,13f6.3,/,8f7.4)
      stop
      end
```

Fig. 1. Fortran program, with parallelization commands for an Intel Paragon.

Fig.2 shows results for $T = 3$, $N_{max} = 5.12 \cdot 10^9$, $M = 1$, minimum age = 8, initial $N = 0.512 \cdot 10^9$, probability of birth 0.11, after 1000 years (not much changed after 10,000 years in smaller runs). We see that the population decreases with increasing age, first slowly, then rapidly. Thus the survival rate is first roughly constant near unity, and then decays rapidly towards about 1/2.

In contrast to many other models[2,6] – but in agreement with Partridge and Barton[7] – the present model allows for a stable population even if all new mutations are negative (as is often assumed in the biological literature[7]). A new mutation affects only the child, not the parent and the other offspring. These other children may thus dominate in the population and keep the average total number of mutations per individuals (number of bits set in 32-bit word) constant, avoiding the usual mutational meltdown[6]. Fig.3 shows results for the case of only bad new mutations (`ieor` replaced by `ior` in the mutations) if we start with no mutations at all and set the birth probability as one; all other parameters are chosen as in Fig.2 except that the system was smaller to fit onto one workstation. Again, population and survival rate decay first slowly, then rapidly with increasing age. If we omit the mutations, on the other hand, all ages have the same survival probability (in general, mutations reduce the population).

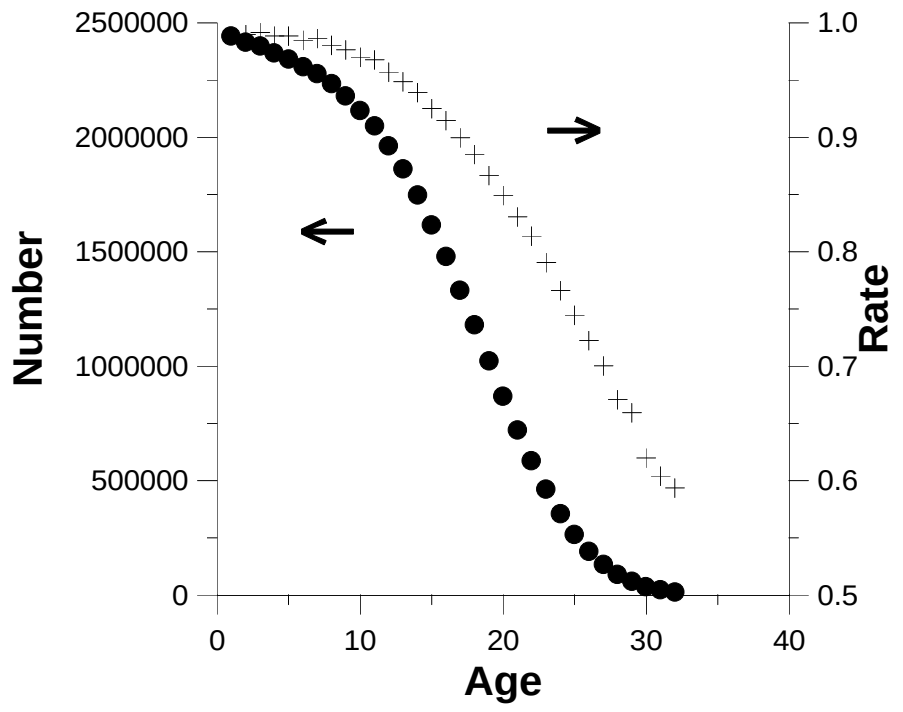

Fig. 2. Decays of population and survival rate with age $t$. In a stationary state like here, the survival rate is the ratio of the population at age $t$ to that at age $t-1$. This simulation with 512 million initial babies took about 12 minutes on an Intel Paragon parallel computer using 128 nodes. (Finite size effects have been observed in the simulation with 200 thousands initial babies).

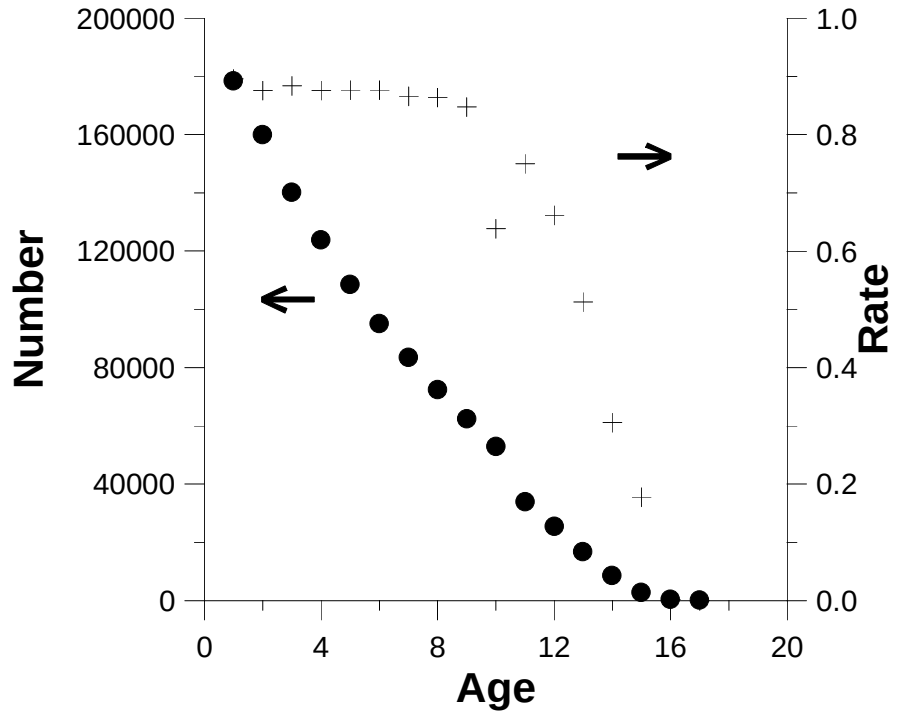

Fig. 3. As fig.2, but without mutations at the beginning and with only negative mutations later. This simulation took about 10 hours on an IBM Powerstation.

Fig.4 shows the changes with the number of iterations ("years") for the simulations of figs.2, using smaller systems. Variation of the above parameters of Figs.2 did not change drastically the qualitative behavior, provided the birth rate was adjusted such that the population did not decay to zero nor increased too rapidly towards the limit imposed by $N_{max}$.

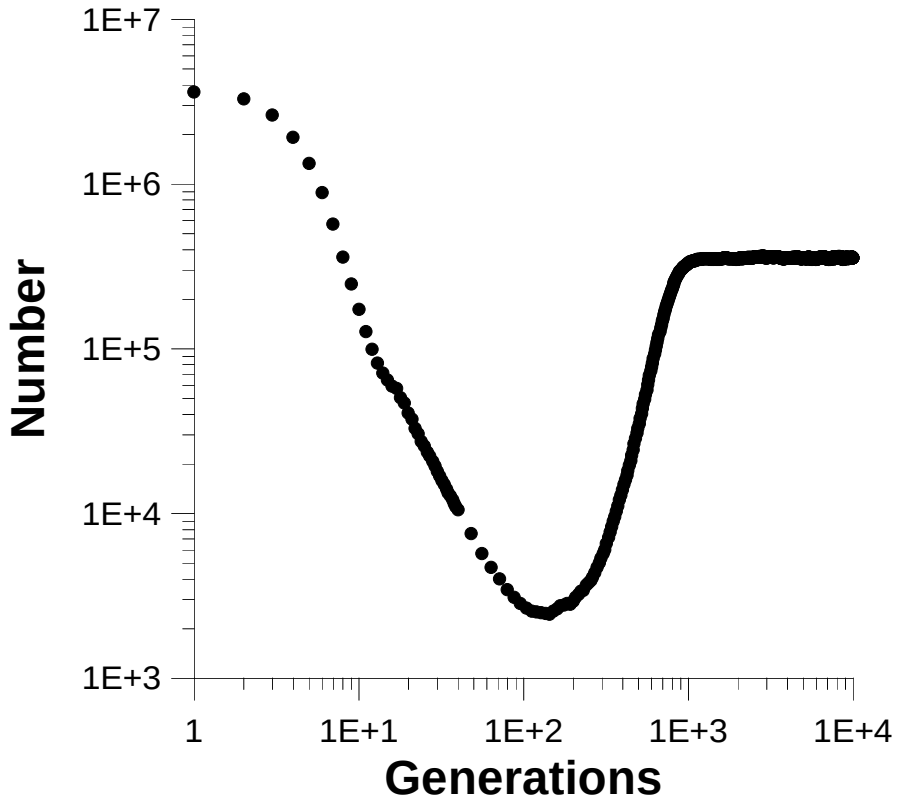

Fig. 4. Development of population with the number of computer interations (corresponding to biological generations or years) showing how a stationary state is established.

Thus this model, similar to other bit-string models in biology [8], gives a simple and efficient way to incorporate many aging intervals into simulations of a life-history model.

## Acknowledgements

We thank the Brazilian Education Ministry for a CAPES travel grant to DS, KFA Julich for time on their Intel Paragon computer and Prof. H.M. Nussenzveig for critical questions. Financial support by Brazilian Agencies CNPq and FINEP is also gratefully acknowledged.